# The Quest for the FFA led to the Expertise Account of its Specialization

Isabel Gauthier, Vanderbilt University

**Abstract**

This article is written in response to Kanwisher's *Progressions* article "The Quest for the FFA and Where It Led" (Kanwisher, 2017). I reflect on the extensive research program dedicated to the study of how and why perceptual expertise explains the many ways that faces are special, a research program that both predates and follows Kanwisher's 1997 landmark article where the fusiform face area (FFA) is named. The expertise accounts suggests that the FFA is an area recruited by expertise individuating objects that are perceptually similar because they share a configuration of parts. While Kanwisher (2017) discussed the expertise account only very briefly and only to dismiss it, there is strong and replicable evidence that responses in the FFA are highly sensitive to experience with non-face objects. I point out that Kanwisher was well positioned to present these findings in their historical context as she participated in the design of the first fMRI study on car and bird expertise, as well as the first replication of this finding. Perhaps most relevant to readers interested in the neural bases of face processing, it is important to distinguish studies that describe the phenomenon of face-selectivity from those that test an explanation for this phenomenon. Kanwisher's claim of a face-dedicated processing module that is not the result of our experience with faces is made without attention to a great deal of expertise research which is directly inconsistent with that claim. The claim also lacks more direct support, as face-selective responses in the visual system are not found in infants and children and face-selective activity in FFA does not appear to be heritable.

In 1997, at Yale, I was preparing to defend my dissertation, entitled "Dissecting face recognition: The role of expertise and level of categorization in object recognition". Four years earlier, I had told my advisor, Mike Tarr, that I wished to learn cognitive neuroscience, then a new and exciting field. I wanted to do brain imaging, study brain-damaged patients and learn all about object recognition so that we might understand how results from these various approaches applied to face recognition. In my memory, and I wish I had it on tape, Mike said "well, I don't really do cognitive neuroscience, but we'll learn it together". By the time I was writing up my thesis, Mike had more than delivered on his promise and we had several wonderful collaborators (Jim Tanaka, Marlene Behrmann, John Gore and Adam Anderson) with whom we started to explore the hypothesis that the acquisition of perceptual expertise, or more specifically the skills we develop in a lifetime of individuating faces, might underlie the specialization of face recognition. While completing my dissertation, I decided to reach out to the person whose own research most clearly ran contrary to my conclusions, Nancy Kanwisher at MIT, to ask whether I could do my postdoctoral research with her. Up to now, I have let the results of this collaboration speak for themselves, but here for the first time I offer some reflections on this experience and what unfolded since.

In a recent *Progressions* article in The Journal, Kanwisher reflected on the impact of a study she published about the same time that I was completing my dissertation. Kanwisher's work, conducted with Josh McDermott and Marvin Chun, used fMRI to localize the fusiform face area (FFA) and characterize its responses. Published in 1997, the paper reporting this research has been highly influential in our field (Kanwisher et al., 1997). The *Progressions* format brings a nice personal dimension to our understanding of neuroscience research, but as the format has a historical flavor, it requires a high bar for accuracy. That is, future students of our field should expect to find in such pieces an accurate account of how influential research programs came to be and have been received. A short paragraph in that article (see Figure 1) addressed the expertise account

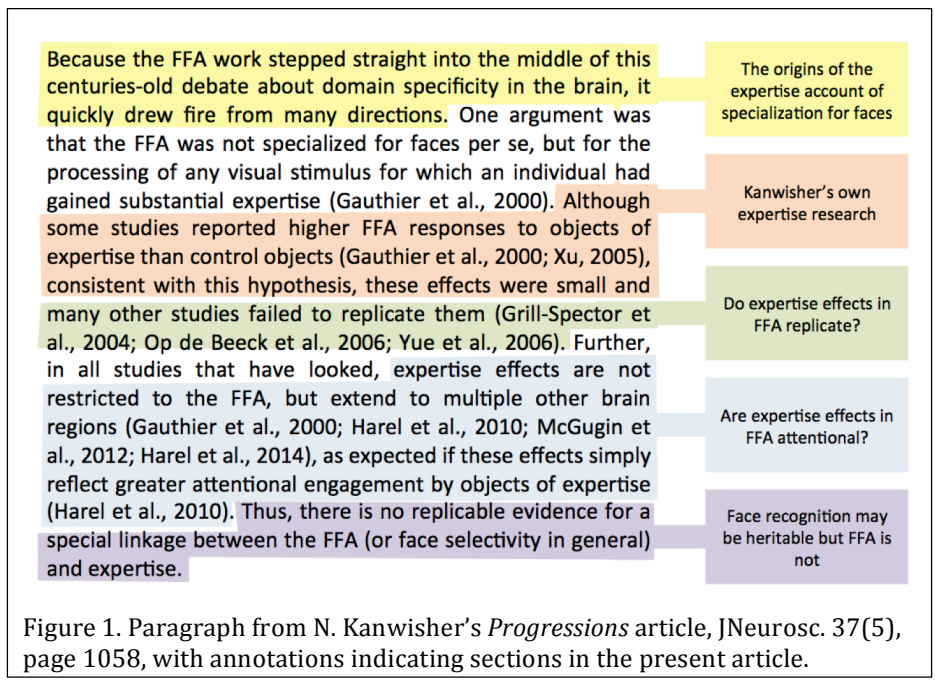

Figure 1. Paragraph from N. Kanwisher's *Progressions* article, JNeurosc. 37(5), page 1058, with annotations indicating sections in the present article.

of the specialization of FFA.

Because my own story intersects with Kanwisher's, I am in a unique position to contribute and add to the record and, because a great deal of my own research has been devoted to the study of expertise, I am also well positioned to offer a defense of the expertise account against Kanwisher's dismissive conclusions. I have organized my perspective into 5 sections, corresponding to the specific points raised in Kanwisher's single paragraph that addresses this alternative, leaving to others any evaluation regarding other claims present in the remainder of Kanwisher's article.

**The origins of the expertise account of specialization for faces**

Kanwisher suggests that the discovery of the FFA was attacked by those disinclined to accept the strong formulation of the modularity hypothesis and this include the alternative account of expertise. However, this perspective does a disservice to a literature devoted to trying to explain the specialization of face recognition that pre-dates Kanwisher's paper. The expertise account of face specialization was not motivated by the demonstration of an FFA (for which no behavioral data were collected), but by behavioral work on the face inversion effect. That is, it was well established in the face processing literature that faces are more difficult to recognize when turned upside down compared other kinds of objects (airplanes, flowers) (Yin, 1969). The Face Inversion Effect was taken as a signature of the special holistic processes reserved for face recognition. However, contrary to this view, Diamond & Carey (1986) demonstrated that dog show judges evince a face-inversion effect that well may be the result of their specific experience. Such findings along with neuropsychological research by Farah and colleagues (Farah et al., 1995) motivated the bulk of my dissertation research based on a set of 3D rendered novel objects we called Greebles. Critically, as with human faces, Greebles share a small number of parts in a common overall spatial arrangement so that individuating them requires encoding the specific shape of their parts and small deviation in their configuration. We trained naïve observers in the lab to individuate these perceptually similar objects, and as observers' response times decreased, we found that they also became very sensitive to the configuration of Greeble parts (Gauthier & Tarr, 1997), just as Jim Tanaka had observed for face stimuli (Tanaka & Farah, 1993; Tanaka and Sengco, 1997).

Kanwisher's 1997 FFA article, as she acknowledges herself, followed in the footsteps of other work describing face-selectivity in the visual system (e.g., Sergent et al., 1992; Haxby et al., 1994; Puce et al., 1996). Again, in her historical account, Kanwisher suggests that her paper "drew fire". Contrary to that characterization, there was already an extant literature discussing explanations for face specialization, with one explanation being that faces become special due to the kind of experience most of us have with them. More to the point, it seems that the empirical findings mainly pointed to a phenomenon that begged explanation. There was nothing in Kanwisher and colleagues' observation of neural face selectivity that helps account for or tests any theory on the origins of such face selectivity.

At the same time, it is worth acknowledging that there has been a tendency towards some scientific backlash whenever a researcher has proffered a singular "spot in the brain" as evidence for a specialized and hard-wired mechanism. To be clear, this fallacy of reasoning, which was perhaps more likely when fMRI was new and shiny, is something that Kanwisher has been careful not to fall prey to in both her original paper and her more recent, historical piece. That being said, Kanwisher rejects, in the strongest possible terms, the expertise hypothesis as a viable alternative account of why we observe face specialization – as such, it is critical to be clear as to what this alternative actually says.

The expertise account of the specialization of regions[1] in the fusiform gyrus for faces does not predict that *every* category of expertise will activate face-selective areas. Rather, it holds that the experience most of us have with faces, individuating a large number of objects with a common configuration of parts, is what leads this area to respond more to faces than to most stimuli. Logically then, visual expertise that differs in critical ways from face processing should not recruit the same brain regions (indeed, expertise with letters or musical notation engages other brain areas, Baker et al., 2007; James et al., 2005; Wong et al, 2010). This framework invites us to study how faces are processed and to devise scenarios where nonface objects are processed in the same manner as faces, in which case, they should recruit the FFA. Looking back, the expertise hypothesis has had significant impact, represented by the extensive body of research that has characterized face-like expertise with many non-face objects, typically associated with holistic processing and sensitivity to configural information (e.g., Boggan, Bartlett, Krawczyk, 2012; Busey & Vanderkolk, 2005; Chua, Richler & Gauthier, 2014). Moreover, extensive practice individuating visually-similar objects leads to face-like processing, whereas extensive practice with other tasks but the same objects does not (e.g., Scott, Tanaka, Sheinberg & Curran, 2008; Wong, Folstein & Gauthier, 2012; Wong Palmeri & Gauthier, 2009).

**Kanwisher's own expertise research**

Remarkably, one of the first studies to examine the neural bases of nonface object processing in experts was the result of a collaboration with Kanwisher. When I was considering what direction to pursue for a post doc, I decided that it would be particularly interesting to work with someone with whom I disagreed. In science, as in politics, we tend to circulate among those who share our opinions. To counter this instinct, I proposed to Kanwisher what others later referred to as an adversarial collaboration[2] (Kahneman, 2003) – realizing that we might never agree on the interpretation, I thought it would be scientifically advantageous to agree on the method for the "next" study that needed to be done. I was bolstered by my own funding from the Natural Sciences and Engineering Research Council of Canada and the generous support of John Gore at Yale to scan volunteers as needed for the project. I approached the collaboration looking for a unique training experience that would hopefully establish a higher bar for this research program. Kanwisher agreed to collaborate with me on this study.

---

[1] The single FFA has since been shown to comprise two distinct functional areas in most brains (Pinsk et al., 2009; Weiner & Grill-Spector, 2010).
[2] I would recommend this experience to all scientists at some point in their training.

Over the course of several months, I visited Kanwisher at MIT so that we could settle on a design for our study. Relevant to this discussion was the publication of an important component of my PhD thesis, an fMRI study of perceptual expertise in which I had trained people in the lab to become visual experts with Greebles. I had found that upright Greebles came to engage the FFA more than upside-down Greebles or familiar objects (Gauthier et al., 1999). Kanwisher thought that these effects could be due to Greebles "looking like faces". It was a fair point (one I later pursued and showed not to be the case, Gauthier, Behrmann & Tarr, 2004) and so we decided together that an appropriate next step would be to test the expertise hypothesis with two categories that were very different geometrically both from each other and from human faces, namely birds and cars. We also thought it important that one category was man-made and the other was not: any result we obtained for both categories could not be easily explained as an artifact of the category itself. After agreeing on an experimental design for our new fMRI study, I recruited about 20 bird and car experts and quantified their individual levels of expertise using behavioral tasks involving bird and car recognition. I then scanned 16 of these experts on a 1.5Tesla scanner at Yale, while Kanwisher scanned another 3 experts on a 3Tesla scanner at MGH in Boston. We found that the FFA in each of our subjects showed neural responses at least twice as much to faces as to objects a clear replication of Kanwisher's earlier results. Critically, this same brain region also showed a stronger response to cars than birds in car experts and a stronger response to birds than cars in bird experts. True to the term "adversarial collaboration," Kanwisher, whose hypothesis was not supported by these results, searched for other plausible alternative explanations and requested a variety of further tests and analyses. I learned a great deal in that process, looking at these results from every possible angle. Perhaps the most important of these analyses is one that to my knowledge was entirely novel for the time, and one that we have replicated many times in the ensuing years (McGugin et al., 2012; 2014; submitted): we examined spatially unsmoothed data and co-registered the different subjects' FFA on their peak of maximum face response. The most face-selective voxel in the FFA also coincided with the peak voxel of the expert response to cars and birds. This and other analyses were published in 2000 in *Nature Neuroscience*, but without Kanwisher as a co-author by her own decision. She did not at the time voice any concern about the soundness of the design we came up with or the analyses that she requested.

Interestingly, Kanwisher requested (and received) permission to use the same stimuli and tasks in future studies. Kanwisher hired a new postdoc, Yaoda Xu, and together they planned to replicate our study. Specifically, Kanwisher and Xu were concerned that the blocked design used in the first bird and car expertise study may have led to more attention to objects of expertise by experts, and so they adopted an event-related design in which they thought these effects were less likely. Summarizing the results of this attempted replication, the Xu paper (Xu, 2005) ends with: "…*the present study reported a strong expertise effect in the right FFA. The results are overall consistent with those of Gauthier et al. (2000) and suggest the involvement of the right FFA in the processing of non-face expertise visual stimulus.*" Kanwisher, again, declined to be an author on the replication study. When, in her *Progressions* piece, Kanwisher suggests that expertise effects in FFA do not replicate, it is historically significant that she designed the original study and the first replication attempt (both my 2000 article and Xu's 2005 paper acknowledge support from her grants). In her latest article, Kanwisher describes the rush she gets when replicating her own results. She speaks to the replication crisis that our field faces and suggests that we need a stronger tradition of replicating our own results before publishing them. This is certainly important, but we also need a tradition of publishing the results of our studies even when they do not support our predictions.

**Do expertise effects in FFA replicate?**

In addition to questioning the replicability of expertise effects, Kanwisher (2017) rejects this account based partly on direct criticisms: that the effect is small, that it may be due to attention, that the effect is distributed rather than locally-constrained to FFA and that it is not replicated in some studies. She also points to findings about face-selectivity that seem to support a face-selective account: the very sharp boundary that goes from face-selective responses in the FFA to non-face selective responses outside the area and the finding that most neurons inside face-selective patches in the monkey, thought to be analogous to human FFA, are highly face-selective. I now evaluate each of these claims and results, showing the criticisms do not hold up to scrutiny. Instead, it may be helpful to realize that studies reporting highly face-selective responses, just as the original report that named the FFA, *describe* a phenomenon, but they do not *explain* it.

Despite evidence to the contrary, Kanwisher suggests that expertise effects are small and do not replicate. First, let us consider the size of expertise effects. While the selectivity for faces versus objects in the FFA is typically measured as a ratio (e.g., (signal for faces – signal for nonface objects)/signal for nonface objects), effects of expertise are measured as a correlation between such a ratio (e.g., (signal for cars vs. signal for non-car objects)/signal for non-car objects) and a behavioral index of expertise. We correlate neural responses with behavioral expertise because the latter can be measured on a continuum. We predict larger selectivity with greater expertise but it is not clear what level of experience with cars would be equivalent to an average level of expertise with faces. Correlations are measures of effect size, and in a 2014 study, to enable a power analysis, we performed a meta-analysis of then-published fMRI studies of expertise with cars (Gauthier et al., 2000, 2005; Xu, 2005; Harel et al., 2010; McGugin et al., 2012). We observed the average expertise effect to be $r = 0.54$, 95% CI [0.40; 0.65]. To detect that effect with a power of 80% at a .05 alpha level, 22 subjects are sufficient. When we ran the new study (McGugin et al., 2014), the maximum car expertise effect we obtained was also $r = .54$, in the right middle FFA (FFA2). It remained significant ($r = .44$) when we restricted the analysis to only face-selective voxels within a very small area (25 mm$^2$) in the center of the FFA.

Second, Kanwisher lists studies in which expertise effects were not found. It is always a challenge to understand null results, but these studies were not without limitations. One measured car expertise behaviorally with modern cars but presented antique cars in the scanner (Grill-Spector et al., 2004). This motivated us to run a behavioral study in which we found that a marker of perceptual expertise for faces, the tendency to process all parts of an object at once (a.k.a. holistic processing) is found in modern car experts when they are tested with modern cars, but not with antique cars. In other words, it is important to test experts with images of stimuli with which they are actually expert. The other two studies (Op de Beeck et al., 2006; Yue et al., 2006) are training studies with artificial objects and small samples of 6 and 9 subjects, respectively. Training effects of expertise are likely to be smaller than those obtained in real world experts, and neither study tested for holistic processing as an indication that the training led to face-like processing. This is a test we have successfully employed in our own training studies, and we find that the activity in the fusiform gyrus of our trained subjects correlates with this behavioral marker (Gauthier & Tarr, 2002; Wong et al., 2009). Aside from studies with birds and cars, and other investigations in my lab demonstrating effects of expertise in the FFA, other studies have found expertise effects for non-face objects in FFA, including for radiological images and for chess pieces configuration (Harley et al., 2009; Bilalić et al., 2011; 2016; Rishi, Tarr & Kingon, 2013).

In addition to being a replicated effect with conventionally medium-to-large effect sizes, and observed in the peak of face-selectivity in the FFA, the expertise effect shows the same sharp boundary that FFA shows for face selectivity. Indeed, face-selective responses drop to zero within a few mm of FFA's standard boundary (Spiridon et al., 2006). In a study where we observed an expertise effect across 25 subjects (McGugin et al., 2012), we looked at the 12 subjects with FFAs extending at least 300 mm$^2$ in flattened cortical space, to consider response selectivity at several concentric, non-overlapping regions of interest. These analyses revealed that the expertise effect was present in a 200 mm$^2$ area, but the correlation was no longer evident outside of this boundary (Figure 2). When Rankin McGugin, in her dissertation, found expertise effects that were highly spatially co-localized with face-selectivity in the fusiform gyrus and were strong even in the most face-selective high-resolution voxels in FFA, she and I immediately thought of the findings by Tsao and colleagues in the monkey (Tsao et al., 2006). Because expertise effects are robust in very small highly face-selective patches, and if almost all the neurons in these patches are face-selective in the human as Tsao found in the monkey, then it follows that the same neurons should participate in the processing of face and non-face expertise. This suggests that face and non-face expertise should compete when they are engaged at the same time. Indeed, such competition has been observed in behavioral (McGugin et al, 2011; McKeeff et al., 2010) ERP (Gauthier et al., 2003; Rossion et al., 2004; Rossion et al., 2007) and fMRI work (McGugin et al., 2015).

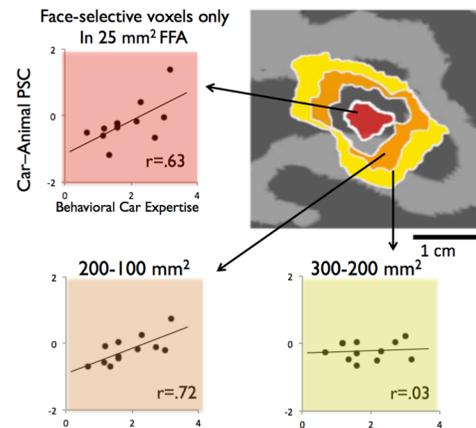

Figure 2. Results adapted from McGugin et al., 2012.

**Are expertise effects in FFA attentional?**

A putative explanation of the expertise effect is that experts attend more to objects of expertise, and indeed attention is known to boost signals at virtually all levels of the visual system (e.g., Wojciulik et al. 1998; Murray and Wojciulik 2004). But it has never been clear to me why this is more of a concern for expertise effects than for face-selectivity (if anything in the visual world draws our attention, it is faces), or how attention was supposed to explain effects that were highly co-localized with hot-spots of face selectivity.

A few specific results speak directly to the question of whether expertise effects in FFA can be explained by attention. Harel et al. (2010) reported that effects of car expertise were distributed in areas outside of FFA and diminished when cars were task irrelevant. We followed up on this in two studies, both manipulating attention and the second replicating the results of the first. In both studies, we found that while expertise effects for cars could be found in several visual areas when the attentional demands were low, when we ramped up these demands using bottom-up and top-down manipulations of attention (e.g., when cars were task-irrelevant) expertise effects were reduced in most areas but remained robust in middle FFA (McGugin et al., 2014a and b). When we used multi-voxel pattern analysis for the first time to look at effects of expertise, we found in that same region patterns of activity for faces and cars that were more similar to one another in experts than in novices (McGugin et al., 2015b). But perhaps the effect that is most obviously impossible to explain by attention is the finding that the regional grey matter thickness of the FFA predicts the ability to recognize cars and other vehicles (McGugin, VanGulick & Gauthier, 2016).

In her article, Kanwisher also suggests that the presence of expertise effects outside of FFA is incompatible with the expertise account. They are not: real-world expertise with cars or chess is likely to affect many different visual processes, only some of which may be supported by FFA. In the same vein, we process faces in many ways, and face selectivity is also found in many areas or patches across the brain including occipital regions (LOC), superior temporal gyrus, temporal pole, and even in frontal regions (e.g., Hoffman and Haxby, 2000; Rossion et al., 2012).

**Face recognition may be heritable but activation in FFA is not**

Kanwisher forcefully dismisses the idea that expertise can explain the origins of the FFA, although later in her *Progressions* article she writes that we do not know whether activation in regions like FFA is determined genetically or by experience. I would argue that the bulk of the evidence since Kanwisher's 1997 paper suggests that activation in FFA is much more determined by experience than by genetic influences. Aside from the strong empirical evidence I reviewed here regarding expertise effects in the FFA, recent work by Kanwisher and her colleagues (Deen et al., 2017) failed to find any visual area in 4-6 month-olds that is more active for faces than objects. Other work has found no stable FFA activity even in 5-8 years old (Golarai et al., 2015; Scherf et al., 2007).

An important distinction between face expertise and activation in the FFA is that there is strong evidence that the behavioral ability to recognize faces (and also cars!) is highly heritable (Wilmer et al, 2010; Shakeshaft & Plomin, 2015), while the activation in the FFA is not. The FFA profile appears to be primarily influenced by experience, even in studies where activation in other visual areas is found to be partly heritable (Pinel et al., 2014). This may explain why it is often difficult to find a relation between the FFA's response to faces and face recognition ability. For instance, we have found that even though the individual differences in neural responses are highly reliable, as are behavioral measures of face recognition, they may not be related, even in the same sample that show an expertise effect for cars (McGugin & Gauthier, 2015). In one very large study (Huang et al., 2014), this relation was found to be much smaller ($r_{294}=0.13$) than the expertise effects around r=.5 that we typically find for non-face objects. It would require 362 subjects to expect replicating this effect with 80% power at an alpha of .05, even with a one-tail test! But in recent work, we reasoned that if we manipulated the level of experience people have with faces of a novel race, then these faces would come to elicit a brain-behavior correlation in FFA comparable in size to the expertise effects we see with non-face objects. This is indeed what we found (McGugin et al., submitted). These results demonstrate, now with faces, the critical role of experience in determining FFA responses.

There is no question that the location of the FFA, highly similar across subjects, is under important constraints (e.g, Hasson et al., 2003). Genetic influences on high-level visual recognition ability could reflect variation in connectivity among several brain areas critical to the acquisition of face (and non-face) recognition ability (see the compromised white matter tracts in individuals with congenital prosopagnosia, which is heritable, Thomas et al. 2009). This acquisition, driven by experience, may fine-tune and optimize neural representations in a number of areas, including the FFA. At least, this is the expertise account of the specialization observed in FFA, and I believe it is at the moment a highly plausible, although most certainly interim, answer to the quest that Kanwisher and her colleagues initiated in 1997.

It would be helpful to fight the impulse to depict evidence of face-selectivity, which is a phenomenon, and the expertise account, which is an hypothesis about the origins of this phenomenon, as being in opposition. We would do better to conduct research that attempts to understand the interaction between innate influences and experience (e.g., Srihasam et al., 2014). And, more than anything, it is important that we keep searching for truth to support our theories instead of using science to support our "truth".

This work was supported by the National Science Foundation (awards #1640681, #1041755. #1534866). I thank the following people for comments: Michael Tarr, Marlene Berhmann, Gary Cottrell, Daniel Bub, Jim Tanaka, Rankin McGugin, Mackenzie Sunday, May Shen and David Sheinberg. The author declares no competing financial interests.

Correspondence should be addressed to Isabel Gauthier, 308 Wilson Hall, Vanderbilt University, Nashville, TN 37240, Isabel.gauthier@vanderbilt.edu

Baker, C. I., Liu, J., Wald, L. L., Kwong, K. K., Benner, T., & Kanwisher, N. (2007). Visual word processing and experiential origins of functional selectivity in human extrastriate cortex. *Proceedings of the National Academy of Sciences*, *104*(21), 9087-9092.

Bilalić, M., Langner, R., Ulrich, R., & Grodd, W. (2011). Many faces of expertise: fusiform face area in chess experts and novices. *Journal of Neuroscience*, *31*(28), 10206-10214.

Bilalić, M. (2016). Revisiting the role of the fusiform face area in expertise. *Journal of Cognitive Neuroscience*, 28, No. 9, Pages: 1345-1357

Boggan, A. L., Bartlett, J. C., & Krawczyk, D. C. (2012). Chess masters show a hallmark of face processing with chess. *Journal of Experimental Psychology: General*, *141*(1), 37.

Bukach, C.M., Phillips, S.W., Gauthier, I. (2010). Limits of generalization between categories and implications for theories of category specificity. *Attention, Performance and Psychophysics*, 1865-1874.

Busey, T. A., & Vanderkolk, J. R. (2005). Behavioral and electrophysiological evidence for configural processing in fingerprint experts. *Vision Research*, *45*(4), 431-448.

Chua, K. W., Richler, J. J., & Gauthier, I. (2014). Becoming a Lunari or Taiyo expert: learned attention to parts drives holistic processing of faces. *Journal of Experimental Psychology: Human Perception and Performance*, *40*(3), 1174.

Deen, B., Richardson, H., Dilks, D. D., Takahashi, A., Keil, B., Wald, L. L., ... & Saxe, R. (2017). Organization of high-level visual cortex in human infants. *Nature Communications*, *8*, 13995.

Diamond, R., & Carey, S. (1986). Why faces are and are not special: an effect of expertise. *Journal of Experimental Psychology: General*, *115*(2), 107-117,

Farah, M. J., Levinson, K. L., & Klein, K. L. (1995). Face perception and within-category discrimination in prosopagnosia. *Neuropsychologia*, *33*(6), 661-674.

Gauthier, I., Behrmann, M., & Tarr, M. J. (2004). Are Greebles like faces? Using the neuropsychological exception to test the rule. *Neuropsychologia*, *42*(14), 1961-1970.

Gauthier, I., Skudlarski, P., Gore, J. C., & Anderson, A. W. (2000). Expertise for cars and birds recruits brain areas involved in face recognition. *Nature Neuroscience*, *3*(2), 191-197.

Gauthier, I., & Tarr, M. J. (1997). Becoming a "Greeble" expert: Exploring mechanisms for face recognition. *Vision Research*, *37*(12), 1673-1682.
10

Tanaka, J. W., & Farah, M. J. (1993). Parts and wholes in face recognition. *The Quarterly Journal of Experimental Psychology*, *46*(2), 225-245.

Tanaka, J. W., & Sengco, J. A. (1997). Features and their configuration in face recognition. *Memory & Cognition*, *25*(5), 583-592.

Tanaka, J. W., & Taylor, M. (1991). Object categories and expertise: Is the basic level in the eye of the beholder? *Cognitive Psychology*, *23*(3), 457-482.

Thomas, C., Avidan, G., Humphreys, K., Jung, K. J., Gao, F., & Behrmann, M. (2009). Reduced structural connectivity in ventral visual cortex in congenital prosopagnosia, Nat Neurosci.12(1):29-31.

Tsao, D. Y., Freiwald, W. A., Tootell, R. B., & Livingstone, M. S. (2006). A cortical region consisting entirely of face-selective cells. *Science*, *311*(5761), 670-674.

Weiner, K. S., & Grill-Spector, K. (2010). Sparsely-distributed organization of face and limb activations in human ventral temporal cortex. *Neuroimage*, *52*(4), 1559-1573.

Wilmer, J. B., Germine, L., Chabris, C. F., Chatterjee, G., Williams, M., Loken, E., ... & Duchaine, B. (2010). Human face recognition ability is specific and highly heritable. *Proceedings of the National Academy of sciences*, *107*(11), 5238-5241.

Wojciulik, E., Kanwisher, N., & Driver, J. (1998). Covert visual attention modulates face-specific activity in the human fusiform gyrus: fMRI study. *Journal of Neurophysiology*, *79*(3), 1574-1578.

Wong, A.C.-N., Palmeri, T.J., Rogers, B.P., Gore, J.C., Gauthier, I. (2009). Beyond shape: How you learn about objects affects how they are represented in visual cortex, *PLoS One*, 2(12), e8405.

Wong, Y. K., & Gauthier, I. (2010). A multimodal neural network recruited by expertise with musical notation. *Journal of Cognitive Neuroscience*, *22*(4), 695-713.

Wong, A. C. N., Palmeri, T. J., & Gauthier, I. (2009). Conditions for facelike expertise with objects: Becoming a Ziggerin expert—but which type?. *Psychological Science*, *20*(9), 1108-1117.

Wong, Y. K., Folstein, J. R., & Gauthier, I. (2012). The nature of experience determines object representations in the visual system. *Journal of Experimental Psychology: General*, *141*(4), 682.

Xu, Y. (2005). Revisiting the role of the fusiform face area in visual expertise. *Cerebral Cortex*, *15*(8), 1234-1242.

Yin, R. K. (1969). Looking at upside-down faces. *Journal of Experimental Psychology*, *81*(1), 141-145.
14